\documentclass[copyright,creativecommons]{eptcs}

\usepackage{amsmath}
\usepackage{iftex}
\usepackage{graphicx}

\usepackage{newunicodechar}
\usepackage{textcomp,gensymb}

\usepackage{xpatch}

\makeatletter
\NewCommandCopy{\renewunicodechar}{\newunicodechar}
\xpatchcmd{\renewunicodechar}{\@ifundefined}{\@gobblethree}{}{}
\makeatother

\newunicodechar{α}{\ensuremath{\alpha}}
\newunicodechar{β}{\ensuremath{\beta}}
\newunicodechar{χ}{\ensuremath{\chi}}
\newunicodechar{ρ}{\ensuremath{\rho}}
\newunicodechar{σ}{\ensuremath{\sigma}}
\newunicodechar{ε}{\ensuremath{\epsilon}}
\newunicodechar{κ}{\ensuremath{\kappa}}
\newunicodechar{τ}{\ensuremath{\tau}}
\newunicodechar{π}{\ensuremath{\pi}}
\newunicodechar{φ}{\ensuremath{\phi}}
\newunicodechar{η}{\ensuremath{\eta}}
\newunicodechar{γ}{\ensuremath{\gamma}}
\newunicodechar{λ}{\ensuremath{\lambda}}
\newunicodechar{ν}{\ensuremath{\nu}}
\newunicodechar{Γ}{\ensuremath{\Gamma}}
\newunicodechar{ω}{\ensuremath{\omega}}
\newunicodechar{Ω}{\ensuremath{\Omega}}
\newunicodechar{Π}{\ensuremath{\Pi}}
\renewunicodechar{°}{\degree}
\newunicodechar{≈}{\ensuremath{\approx}}
\newunicodechar{≅}{\ensuremath{\approxeq}}
\newunicodechar{≡}{\ensuremath{\equiv}}
\newunicodechar{≤}{\ensuremath{\le}}
\newunicodechar{≥}{\ensuremath{\ge}}
\newunicodechar{≠}{\ensuremath{\ne}}
\renewunicodechar{²}{\ensuremath{^2}}
\renewunicodechar{³}{\ensuremath{^3}}
\newunicodechar{∂}{\ensuremath{\partial}}
\newunicodechar{∇}{\ensuremath{\nabla}}
\renewunicodechar{⟨}{\ensuremath{\langle}}
\renewunicodechar{⟩}{\ensuremath{\rangle}}
\renewunicodechar{⋅}{\ensuremath{\cdot}}
\renewunicodechar{→}{\ensuremath{\rightarrow}}
\renewunicodechar{←}{\ensuremath{\leftarrow}}

\newcommand{\pp}[3][]{{\frac{\partial^{#1} #2}{\partial #3^{#1}}}}

\makeatletter
\let\oldabs\abs
\def\abs{\@ifstar{\oldabs}{\oldabs*}}
\let\oldnorm\norm
\def\norm{\@ifstar{\oldnorm}{\oldnorm*}}
\makeatother

\newcommand{\mO}{\textnormal{O}}

\mathchardef\mhyph="2D

\usepackage{siunitx}
\ifpdf
  \usepackage{underscore}         
  \usepackage[T1]{fontenc}        
  \usepackage{microtype}
\else
  \usepackage{breakurl}           
\fi

\title{Property Testing for Ocean Models: Can We Specify It? (Invited Talk)}
\author{Deepak A. Cherian
\institute{Earthmover PBC}
\email{deepak@cherian.net}
}
\def\titlerunning{Property testing ocean models}
\def\authorrunning{D.A. Cherian}
\begin{document}

\maketitle

\begin{abstract}
I take inspiration from the property-testing literature, particularly the work of Prof.\ John Hughes \cite{hughesHowSpecifyIt2020}, and explore how such ideas might be applied to numerical models of the ocean.
Specifically, I ask whether geophysical fluid dynamics (GFD) theory, expressed as property tests, might be used to address the oracle problem of testing the correctness of ocean models.
I propose that a number of simple idealized GFD problems can be framed as property tests.
These examples clearly illustrate how physics naturally lends itself to specifying property tests.
Which of these proposed tests might be most feasible and useful, remains to be seen.
\end{abstract}

\section{Introduction}

\subsection{Background}

\begin{figure}
    \centering
    \includegraphics[width=0.4\linewidth]{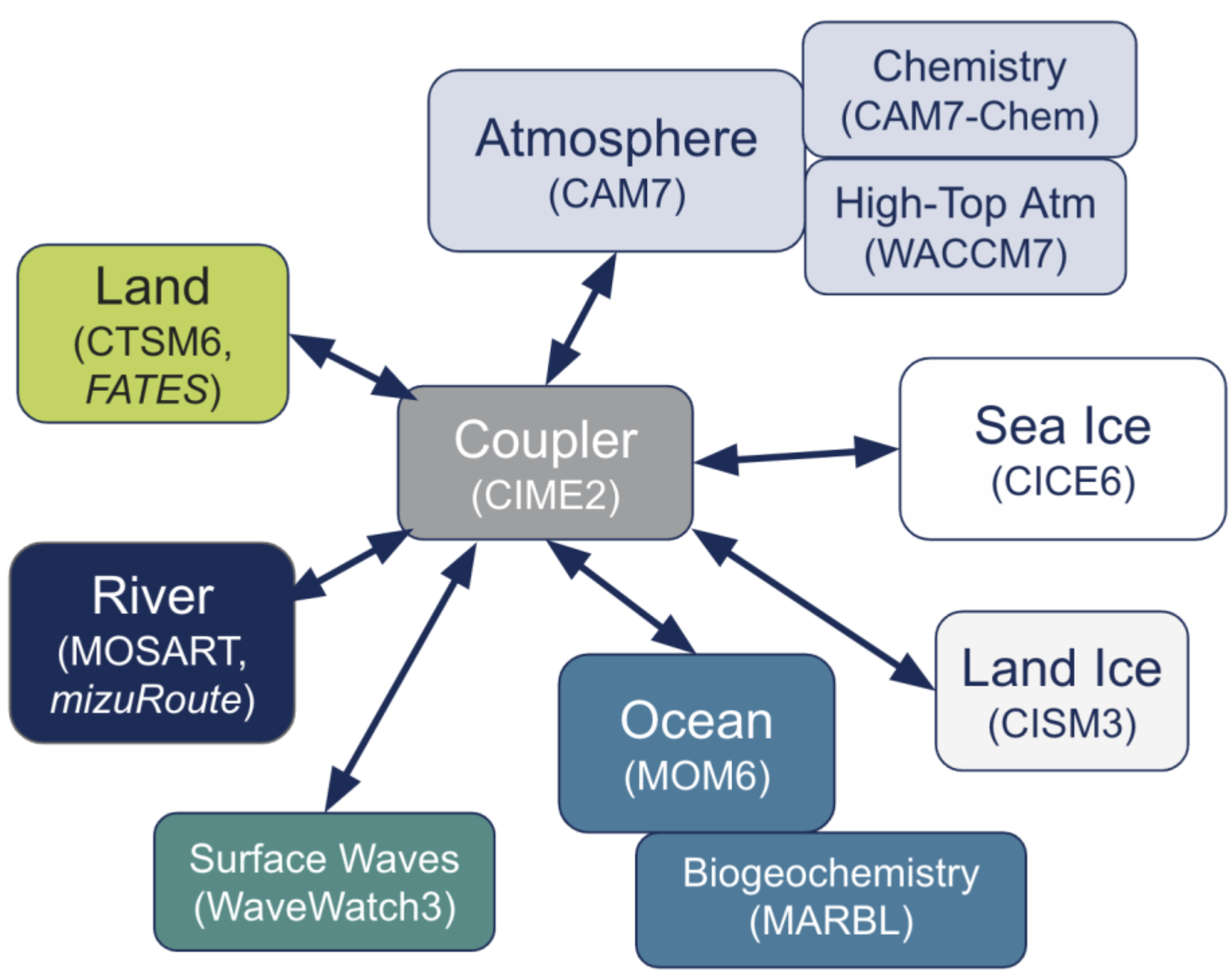}
    \caption{Schematic of the Community Earth System Model version 3 (CESM3) \cite{lawrenceCommunityEarthSystem}.}
    \label{fig:cesm}
\end{figure}
\begin{figure}
    \centering
    \includegraphics[width=0.9\linewidth]{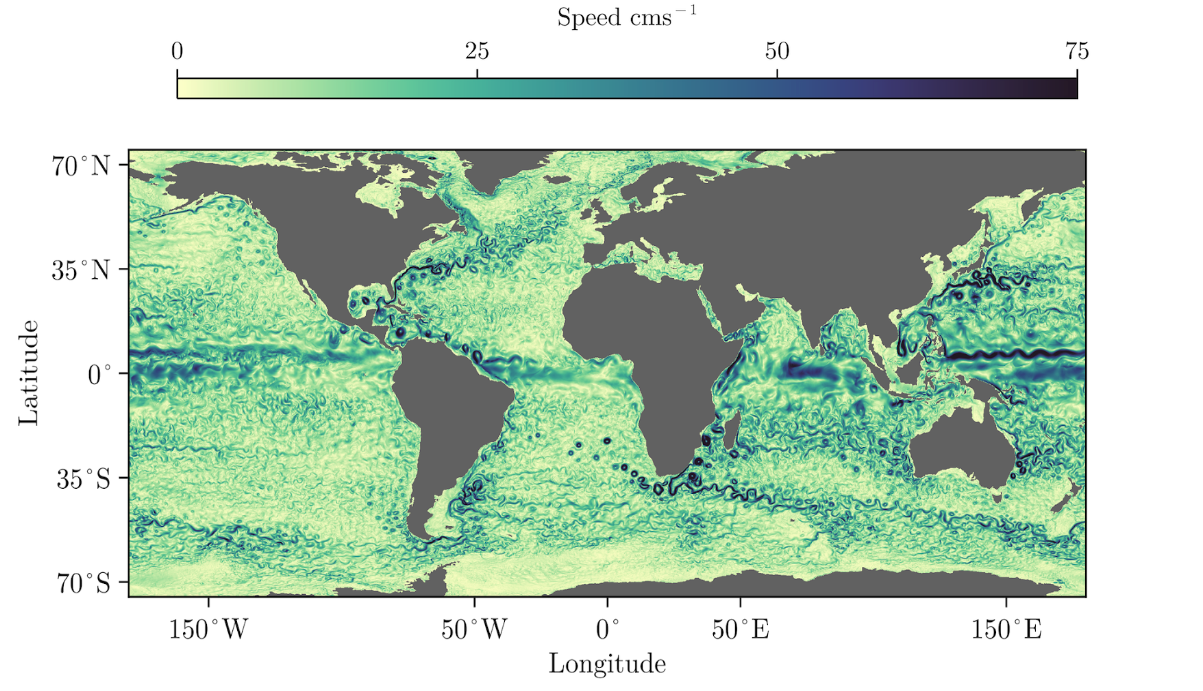}
    \caption{Sea surface speed from a global 1/12° simulation using the Oceananigans model \cite{Silvestri2024blog}. Note several properties of the ocean circulation: (a) faster currents at western continental boundaries of ocean basins, (b) the east-west orientation of flows in the Tropics away from continental boundaries, and (c) the presence of coherent vortices (circular features).}
    \label{fig:ssspd}
\end{figure}
\begin{figure}
    \centering
    \includegraphics[width=0.9\linewidth]{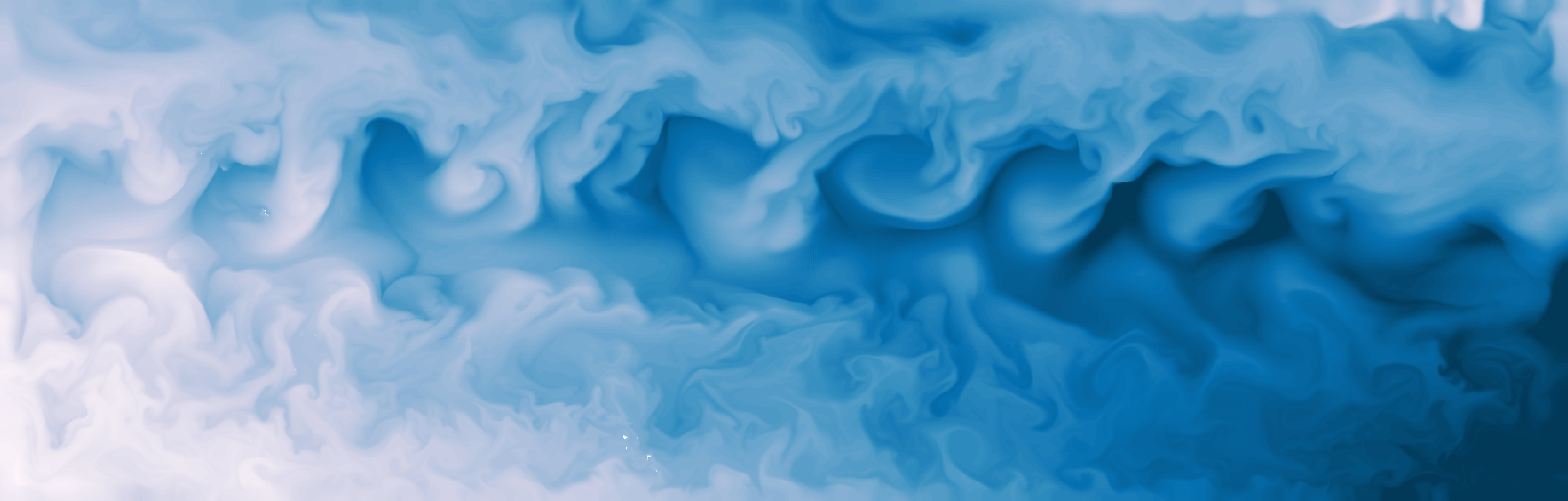}
    \caption{Sea surface temperature in the Tropical Pacific from a regional MIT General Circulation Model simulation using a $1/20°\approx\SI{5}{\km}$ spacing \cite{Cherian2021}. Darker colors indicate colder water. The full domain spans 20 degrees of latitude and 80 degrees of longitude. The complex and turbulent nature of the flow is evident.}
    \label{fig:sst}
\end{figure}
Simulated possible projections of the Earth's climate by Earth System Models are an important tool in understanding, adapting to, and mitigating the effects of climate change.
Such models bring together a number of individual component models (e.g. the atmosphere, ocean, land, rivers, and more) coupled to each other through modeled interactions at the interfaces between these components (for example, the sea surface).
The complexity of each individual component is vast, the complexity of a full Earth System Model is staggering (Figure \ref{fig:cesm}).

Consider the ocean.
The ocean is a thin shell of fluid on the Earth\footnote{compare its depth of $\sim$\SI{5}{\km} to the radius of the earth ($\sim$\SI{6400}{\km})}.
The ocean is ``stratified'', meaning the density of seawater varies with depth so that lighter fluid overlays denser fluid.
Commonly, warm and fresh (low salinity) water overlays colder, saltier water.
The equations that describe the motion of the ocean are the Navier-Stokes equations specialized for rotating fluids.
These equations express the conservation of momentum in three dimensions (\(x\), \(y\), and \(z\)) for three velocity components (\(u\), \(v\), and \(w\)); conservation of mass; conservation of heat and salt; and a thermodynamic ``equation of state'' that relates density $ρ$ to temperature $T$ and salinity $S$.
A similar set of equations can be written down for the atmosphere.
The existence of unique solutions for this set of equations is unproven \cite{Fefferman2006}.
Solutions for the case of geophysical fluids are further complicated by the presence of complex boundary conditions.
For the ocean this includes the oceanic boundary with the landmass at its side, bottom topography, and the surface boundary with the atmosphere across which both kinetic energy and heat flow.
Flows in both the ocean and atmosphere are complex and turbulent.
The range of motions in the ocean range from the ``basin-scale'', approximately \SI{5000}{\km}, to the smallest scales where molecular viscosity and diffusivity take charge, approximately a micrometer.
Given the computational expense, global-scale models tend to solve the equations of motion of grids with spacing \mO(\SI{10}{\km})--\mO(\SI{100}{\km}) or so.
The consequences of physics at scales smaller than the grid spacing is represented using crude approximate models called ``parameterizations''.
Common parameterizations model the effect of turbulence, waves, clouds and other phenomena too small to be simulated directly.
The complexity of the system is immense, and the computational expense of solving these equations is large.
Figures \ref{fig:ssspd} and \ref{fig:sst} illustrate the complexity of modeled ocean flows using surface speed and surface temperature.

Similar observations of complexity extend to all other components of an Earth System Model (Figure \ref{fig:cesm}).
Projections constructed using these models form part of the scientific backing for the reports of the Intergovernmental Panel on Climate Change (IPCC).
These models also enable experimentation and the creation of new knowledge by allowing scientists to tease apart relationships that are too complex to be gleaned by analytical study of the equation set.
It is essential that these models be as correct as we can make them.

\subsection{Testing climate models}
The testing of climate model components suffers from the ``oracle'' problem --- given a numerically constructed solution for the evolution of a system from an initial state, there is no ``oracle'' to tell us whether the solution is correct.
When configured to match the real ocean, we might use observations to judge the quality of an integration, but this approach suffers from the problem of ``compensating errors'' which may hide real bugs.
Unsurprisingly, testing methodologies for ocean/atmosphere models therefore rely heavily on regression tests, and comparison to reference solutions.
Regression testing is treated strictly, with developers emphasizing bitwise reproducibility when comparing to reference solutions or solutions from an earlier version of the model.
Reference solutions to a somewhat arbitrary selection of initial value problems for the atmosphere have been published \cite{galewskyInitialvalueProblemTesting2004,heldProposalIntercomparisonDynamical1994,polvaniNumericallyConvergedSolutions2004,williamsonStandardTestSet1992}.
Interestingly \cite{williamsonStandardTestSet1992} provide error metrics and recommend constructing specific plots for visual inspection as a testing strategy.

A quick survey of ocean modeling systems highlights the similarity in approach to testing.
The MIT General Circulation Model (MITgcm) project provides reproducible test examples with reference output for comparison \cite{mitgcmtests}.
Similarly the Regional Ocean Modeling System (ROMS; \cite{ROMStestcases}), the Coastal and Regional Ocean Community Model (CROCO; \cite{CROCOtestcases}), and the Modular Ocean Model (MOM6, \cite{MOM6tests}) all take a similar approach.
Intercomparison with models is another approach to correctness; for example \cite{Huang2008} report comparing the Finite-Volume Coastal Ocean Model (FVCOM) against ROMS.
Other testing approaches common to many modeling systems include testing that output from a serial and parallel integration are the same, and that a simulation with no restarts is identical to one with a restart in the middle\footnote{Earth System Models can write ``perfect restart'' files that contain all the state necessary to restart a simulation with no changes in the solution. This capability is necessary for conducting long simulations that may take days of wallclock runtime.}.

The developers of the MOM6 system employ two unique testing strategies --- termed ``novel tests'' \cite{MOM6noveltests}.
One is to assert dimensional consistency of the code, after having hand-coded physical units for terms in the codebase.
The second is to test that simulations are invariant to a rotation of the domain by some number of 90° turns.
These ``novel tests'' are property tests in all but name.

More recently, lightweight formal methods are beginning to be applied to sub-components of these models.
For example, Altuntas and Baugh \cite{altuntasHybridTheoremProving2018} use a hybrid theorem prover to test for bugs in a widely used ocean model parameterization (K-Profile Parameterization; KPP, \cite{Large1994}).
Altuntas et al (this volume) systematically argue for greater use of formal methods across all sub-components of a climate model systems.

\subsection{Generative property-based testing}
One solution to the ``oracle'' problem is to use generative property-based testing as popularized by the Haskell library QuickCheck \cite{Claessen2000}.
In this approach, the programmer specifies properties of the System Under Test, and designs a generator for valid inputs to the test.
Outputs derived from a wide range of generated inputs are then tested for correctness using these properties.
Such property tests are increasingly popular, particularly in the form of ``metamorphic testing'', where a ``metamorphic relation'' describes the relationship between outputs generated from related inputs (for example, \cite{Chen2009}).
In the scientific Python ecosystem, the Hypothesis library \cite{maciverHypothesisNewApproach2019} is increasingly used for testing software used for scientific analysis, including that of climate model output.
For example, the author has helped build nascent suites of property and stateful tests in the Xarray \cite{xarrayproptest} and Zarr \cite{zarrproptest} scientific Python projects.

This paper is a speculative exploration of how we might apply property-based testing ideas to the testing ocean modeling systems, guided by experience building such test suites for Xarray and Zarr. 
We are led to ask: ``what are the properties of the ocean circulation?'' and are inevitably led to the field of ``geophysical fluid dynamics''.

\subsection{Geophysical Fluid Dynamics (GFD)}
The complexity of geophysical fluid systems does not only affect numerical approaches.
It took nearly 300 years of analytic effort to write down the modern formulation of the equations of motion \cite{Darrigol2005}.
Over that time and since then, much work has been done to understand these equations in much simpler limits that are amenable to theoretical exploration.
The broad field of geophysical fluid dynamics (GFD; \cite{vallisGeophysicalFluidDynamics2016}) is the study of fluid motion governed by these equations (specialized to the Earth or other planets), with the goal of elucidating fundamental physical principles and relationships.
Geophysical fluid dynamicists have studied flows in a hierarchy of complexity from the shallow water equations (constant density fluid), to quasi-geostrophic flow (slow flows in large domains with relatively large density variations in the vertical), to primitive equation models such as those used to construct the solutions in Figures \ref{fig:ssspd} and \ref{fig:sst}.
Solutions to equations with varying levels of approximation, and under specific cases, are available in the textbooks \cite{Gill1982,Pedlosky1987,Vallis2017}.
Below, I use these simplified cases and explore how they may be framed as property tests.
These simplified GFD problems use simpler domains (e.g. a flat bottom or a square basin) and commonly apply approximations to the representation of the effect of the Earth's rotation, and the complexity of vertical variations in density.
The former is a recurring theme in the rest of the manuscript, so I discuss it briefly.

\subsubsection{Representing a rotating planet}
\label{sec:rotation}
The influence of the Earth's rotation in our rotating frame of reference is expressed through the Coriolis force terms $-fv$ and $fu$ in the $x$ and $y$ momentum equations, where $f = 2Ω\sinφ$ [\SI{}{\per\second}] where $Ω$ is the angular velocity of the Earth and $φ$ is the latitude \cite{Price2006coriolis}.
Thus the effect of Earth's rotation on fluid flow varies with latitude, with its magnitude being negligible at the equator ($φ=0$), and maximum at the poles ($φ=90°$).
A truncated Taylor series expansion of $\sin φ$ is used in simpler models that cover limited spans in latitude $f = f_0 + β y$.
Two common limits are the $f$-plane where $β=0$, and the $β$-plane where $β ≠ 0$.
On the $f$ plane there is no difference between the latitude and longitude directions.
On the $β$ plane, oceanic motions are preferentially oriented along lines of constant latitude (``zonal''), and there is an east-west asymmetry due to the physics of a new class of wave --- the Rossby wave.
Both of these \emph{properties} will be used later.


\section{Dynamical property tests for ocean models}

The most evident property of the ocean circulation is the east-west asymmetry: flows at western boundaries are more energetic (Figure \ref{fig:ssspd}). 
Yet testing the ability to reproduce this property simply tests that the code is able to simulate a rotating spherical planet.
Surely this property is too simple! How might we do better?

Prof. John Hughes lays out a guide with 5 guiding suggestions for thinking about property tests in a paper \cite{hughesHowSpecifyIt2020} and in following talks (for example, \cite{Hughes2020lambda})\footnote{I acknowledge inspiration from at least two other authors: Hillel Wayne \cite{Wayne2019} and Sean Wlaschin \cite{wlaschinPropertyBasedTesting2014}.}.
Here I follow those suggestions and ask how might we write property tests for ocean models.
As an exercise I constrain myself to writing out dynamical properties for the \emph{whole} system, albeit in simple limits, rather than focus on individual terms or subroutines that comprise an ocean model.
This is not to suggest that such tests are most likely to be useful at finding bugs.
Instead this constraint simply makes the exercise more fun and interesting (to me).
Though the discussion is centered on oceanography, nearly all of it can be easily translated to atmospheric and other geophysical fluid models.

In writing out these ideas I find that
\begin{enumerate}
\item many of these ideas are already used as either example problems or regression tests in the current generation of ocean models; and that 
\item the physics of the underlying system clearly yields several dynamical properties. Yet it is not obvious which might be most useful in finding bugs.
\end{enumerate}

Now I address each of Hughes' suggestions in turn.
For each idea, I describe a simple property test for the problem of sorting a list of integers that helps motivate the thinking underlying a proposed oceanographic property test.

\subsection{Is there an invariant?}
\emph{Example}: Sorting is idempotent so sorting a sorted list yields the original sorted list.

\subsubsection{Conservation of properties}
\label{sec:conservation}

Physical systems naturally have integral invariants namely the conservation of mass (or volume), energy, angular momentum, and ``potential vorticity'' \cite{mcintyrepv}.
Given that, one might configure a set of arbitrary initial conditions for the ocean state, integrate the model forward for $n$ timesteps, with \(n\) chosen to limit computational expense.
At the end of the integration, conservation principles must be satisfied to the level expected from the numerics of the model.
However totally arbitrary initial conditions are not possible, they will need to be carefully designed to prevent model blowup.
It seems more useful to test such invariants in the context of other more physically motivated tests below.

\subsubsection{Symmetries}
\label{sec:symmetries}

The Navier-Stokes equation set has underlying symmetries.
The standard example is ``Galilean invariance'' which means the solution is invariant to switching to a reference frame with a constant translating velocity.
Other symmetry groups include rotational symmetry (currently used in the MOM6 ``novel tests''), and a more general ``scale invariance'' \cite[their equations 1.9--1.12]{Majda2001}.
These invariants are all applicable to the specific solutions used in the tests described below.

\subsubsection{Balanced flows remain balanced}
\label{sec:balance}

A more dynamically motivated invariant, in the spirit of this article, is the idea that \emph{``balanced flows must remain balanced''}.

When a flow is \emph{balanced} this means that the velocity and pressure fields are related functionally, and that one is in principle derivable from the other \cite{mcintyrebalanced}.
Crucially, a \emph{lack of balance} is characterized by freely propagating waves (\emph{transients}) that complicated the solution dearly.
For ease of comparison, we will need to design tests that result in either no transients, or known transients.
In this section, we initialized the model with a balanced flow. 
If truly balanced to the level that the numerics allow, the transients must be negligible.
Further we can test our specification of balanced flows using property tests too!
The classic roundtrip property test applies: by definition, transforming from velocity to pressure and back, or pressure to velocity and back, must yield the initial field.

The simplest example here is ``geostrophic balance'' --- the balance between the pressure gradient force and the Coriolis force:
\begin{equation}
fu = -\frac{1}{ρ}\pp py; \qquad fv = \frac{1}{ρ}\pp px
\end{equation}
For arbitrary choices of \(f\) and \(u,v\) (say), there exists a pressure field $p$ that balances that \(u,v\).
If initialized with such a balanced field, the model must maintain this balance.
Classic choices for \(v\), \(p\) are circular flows (vortices) or straight flows (jets).
When $f$ is a constant, there is no dynamical difference between the \(x\) and \(y\) directions so the model must be able to maintain jets oriented at arbitrary angles.
Other formulations of balanced flows add density stratification --- ``thermal wind balance''; and nonlinear advection --- ``cyclostrophic balance''. 

For more complexity, we turn to the idea of \emph{balanced waves} to bring in the time derivative term.
Many types of balanced waves exist in all approximations to the governing equations, all of these would make good tests.
Examples include surface gravity waves at the air-sea interface familiar from visits to the beach, internal gravity waves (similar to surface waves but propagate at internal density interfaces within the fluid), and much larger and slower Rossby waves that can exist in rotating fluid systems.
The waves that exist for a system are described by the \emph{dispersion relation} which relates the wavenumbers and frequencies of waves that can exist given a chosen domain.
For example, internal waves in a rotating stratified fluid satisfy the dispersion relation
\begin{equation}
\label{eq:intwave}
ω^2 = \frac{f^2m^2 + N^2(k^2 + l^2)}{k^2 + l^2 + m^2}
\end{equation}
where $ω$ is the frequency and $(k,l,m)$ are the three components of the wavenumber vector.
Exact solutions can be written depending on the chosen analytic form of $N$ and the boundary conditions.
If initialized with a wave field with the right analytic form and for wavenumbers and frequencies that satisfy the dispersion relation, the model must maintain this wave field forever.

For a more complicated balance, consider modons \cite{Stern1975}.
Modons are propagating solitary-wave like solutions that translate with constant velocity, while preserving their shape.
Their dynamics are strongly non-linear.
For a test, one might assert that the anomalies in a modon's shape are minimal in a frame of reference translating with the expected translation speed \footnote{This test is used in the ROMS suite of tests \cite{Huang2008}.}.
Though complex, specific analytic solutions are available (for example see Boyd \cite{Boyd2018}, their Section 16.14; eqns 16.54--16.56).
Again, if initialized properly the model must maintain and propagate the modon without dispersion.

\subsubsection{Using invariants for validity testing}

Hughes \cite{hughesHowSpecifyIt2020} mentions invariants in the context of testing the code that generates input for property tests.
This idea is particularly applicable to the balanced flow invariant.
For example, we can test our specification of geostrophic balance using additional property tests.
First, geostrophic flows are non-divergent so that $u_x + v_y = 0$.
Second, the geostrophic velocity vector $(u,v)$ is at a right angle to pressure gradient vector $(∂p/∂x, ∂p/∂y)$.
The latter property is evocatively taught to students: flow around a low (high) pressure is counterclockwise (clockwise) in the Northern Hemisphere; the directions are reversed in the Southern Hemisphere.
The symmetry properties (Section \ref{sec:symmetries}) are potentially quite useful in this context too, since they apply to arbitrary flows.

\subsection{What is the postcondition?}
\label{sec:postcondition}
To quote Hughes \cite{hughesHowSpecifyIt2020}: ``Postconditions relate return values to arguments of a single call. \ldots{} We can finesse this problem using a very powerful and general idea, that of constructing a test case whose outcome is easy to predict.'' \\
\emph{Example:} If we insert the minimum representable integer \(k\) into a list of integers and sort that list, \(k\) must be the first element of the sorted list.

What can we insert in to the ocean?
Energy, through either heat (potential energy) or momentum (kinetic energy).
The ocean \emph{adjusts} to imparted energy through transients (usually radiating waves) and eventually evolves to a balanced state, usually geostrophic balance, assuming the system is not unstable.
Nominally a property test might assert that stable systems will \emph{eventually} reach a balanced state.
Doing so would require very long integrations, greatly limiting the usefulness of such a test.
In some rare cases, exact analytic solutions with transients may be calculated \cite[their Section 7.3; pg. 196]{Gill1982}.
For a more general approach, I propose crafting anomalies that result in transients of known properties.

\subsubsection{Resonant frequencies resonate}
\label{sec:resonance}
The balanced wave idea can be extended to include atmospheric forcing of an ocean basin.
For simple basin configurations, we can analytically determine the dispersion relation of a whole host of balanced waves.
These ``free modes'' of the system can be freely excited by atmospheric forcing of the same wavenumber and frequency --- called ``resonance''.
The ``postcondition'' of inserting energy at resonant frequencies is energetic motion.
At these frequencies, energy is very efficiently imparted in to the ocean.
Conversely, no amount of energy input will excite energetic motions at frequencies or wavenumbers that are not on a dispersion relation.
The ``postcondition'' of inserting energy at non-resonsant frequencies is damped motion that dies quickly.
Both are testable properties. 

\subsubsection{Symmetry yields asymmetry}
\label{sec:asymmetry}

A non-zero β parameter introduces a fundamental east-west asymmetry.
In the mid-latitudes (poleward of $\approx20°$), the westward propagating Rossby wave modes are faster and have larger wavelengths.
Eastward propagating modes are smaller wavelength and are dissipated by friction faster.
This fundamental east-west asymmetry of response yields a useful property. 
Inserting energy at the western boundary of a square basin, should yield a response of eastward-propagating waves with a much shorter longitudinal length scale than inserting energy at the eastern boundary.
The latter should yield a response of westward propagating waves of comparably larger length scale.

\subsection{Metamorphic relations}
\label{sec:metamorphic}

To quote Hughes \cite{hughesHowSpecifyIt2020}: ``Related calls return related results.''\\
\emph{Example:} sorting a sorted list must yield the same list, or sorting any permutation of a list must yield the same result.

Metamorphic property relates the outcomes of two related experiments.
In essence, this is the goal of GFD --- to tease out parametric dependencies or functional relationships, or identifying controlling parameters \cite{Vallis2017}.
In simple systems, a handful of parameters (e.g. stratification \(N\), \(β\), domain size) determine fundamental scales of balanced flows, and the predicted relationships can be easily tested (see equation \ref{eq:intwave} for an example).
For example, a doubling of $β$ should double the phase speed of a Rossby wave, or halve the time taken to receive a signal from the eastern boundary.
Similar metamorphic relations can be constructed for all the proposed property tests earlier.

Metamorphic relations might offer a way to handle complex transient fields.
Consider a rare case where the exact time-dependent analytic solution is known.
Gill \cite[their Section 7.3]{Gill1982} discuss the analytic solution to the adjustment of a bump in sea surface height for a rotating constant density fluid.
The solution has two parameters $P≡η_0\sqrt{g/H}$ and ``deformation radius'' $L_D ≡ \sqrt{gH}/f$, where $g$ is the acceleration due to gravity, $H$ is the water depth, $f$ is the rotation rate, and $η_0$ is the initial height of the bump.
Again metamorphic tests can be designed by co-varying $g,H,f,η_0$ so that $P$ and/or $L_D$ are kept constant.
Though not too useful when the analytic solution is known, we may apply the metamorphic relation when the analytics solution is unknown but the controlling parameters are known (e.g. the deformation radius $L_D$).
This proposal will need extensive study but seems plausible.
As an example, for the asymmetric response test (Section \ref{sec:asymmetry}) we might insert wind forcing in the center of a square basin, and integrate for $β>0$; and for $β<0$.
These solutions should be anti-symmetric, and presumably easily comparable.

Finally, like all branches of fluid physics, GFD relies greatly on the concept  of ``dynamical similarity'' --- the idea that the core physics of a problem is determined by a handful of non-dimensional parameters.
For example, there is a flow regime where the response of flow of speed $U$ over sinusoidal topography of wavelength $k$ on the $β$-plane is governed by the non-dimensional parameter $λ = Uk/β$ \cite[pg. 525]{Gill1982}.
If $U$, $k$, and $β$ were all individually rescaled so that $λ$ remained constant, the solution for all these possible combinations of parameters is the same after appropriate rescaling.
This statement is a metamorphic test derived from physics.


\subsection{Inductive properties}
\emph{Example:} All subsets of a sorted list are sorted.

It is not apparent to me how the idea of induction might apply to a fluids system.

\subsection{Model-based properties}
Model-based tests compare the system under test to a much simpler reference implementation.

The previous sections are suffused with model-based properties, unsurprising given that our test ideas are derived from theoretical studies of a simplified model of the full equation set.
In some cases, these simplifications can be relaxed through the use of simpler numerical models.
For example, the dispersion relations for waves in systems with complex bottom topography and density variations can be hard to determine analytically; but possible to derive numerically \cite{brinkcode}.
Such numerically determined dispersion relations may be used to construct property tests for the more complex ocean model.
Similarly, modon and other solitary wave solutions are hard to construct analytically but may be determined from simpler numerical models \cite{Crowe2024}.

\section{Discussion}

Fluid physics naturally yields a large number of dynamical properties that may be recast in a form suitable for generative property-based testing.
In a way, these proposed tests are integration tests that exercise the ability of a model to reproduce the expected physics of a simplified system.
There is no new physics here.
The core innovation here is to recast known ocean physics in the form of property tests, guided by the expertise of the computer science community.
That many of the proposed testing ideas are already implemented as example problems or regression test cases in many ocean modeling systems suggests that this approach may be profitable.

It is quite apparent that the core physics challenge here is the handling of transient motions.
Either one uses flows that are balanced and hence excite no transient motions (Section \ref{sec:balance}), or one specifies initial conditions that excite transient fields of known properties (Section \ref{sec:postcondition}). 
Metamorphic relations, expressed through the dynamical similarity of fluid flows, can also offer a way out of the difficulties posed by transients (Section \ref{sec:metamorphic}). More exploration of this idea is necessary.

It is possible to specify property tests at the level of individual terms in the equations or for specific physical processes. 
As an example, consider the submodule that handles advecting (or moving) a ``tracer variable'' such as dye, temperature, or salinity. 
In a domain that is both periodic in $x$ and $y$, there is a background velocity $(u,v)$ for which a patch of dye will return to its starting location after $N$ time steps. 
The challenge is whether the codebase is modular enough to allow submodules to be tested individually.

Two major unaddressed challenges here are:
\begin{enumerate}
    \item Can we design appropriate shrinking strategies? Shrinking is the idea that generated test cases can be shrunk to yield simpler and more minimal test cases. Shrinkers should be designed such that the simplifying physics assumptions underlying the test are not broken. 

    \item Can we limit the computational expense of the proposed tests to make them feasible?
\end{enumerate}
A concrete implementation of the proposed ideas will be necessary to answer these two questions, and to provide guidance on which property tests are most useful at finding bugs.

I end with a final quote from Hughes \cite{hughesHowSpecifyIt2020}: ``every bug is found by at least one postcondition, metamorphic property, and model-based property.''
Luckily we have many ideas that lie in those three categories!

\nocite{*}
\bibliographystyle{eptcs}
\bibliography{references}
\end{document}